# ATTENTION: *ATT*ack*E*r Traceback using MAC Layer Ab*N*ormality Detec*TION*

**Yongjin Kim, Ahmed Helmy**
*Electrical Engineering Dept. – Systems*
*University of Southern California, California, U.S.A.*
*yongjkim@usc.edu, helmy@usc.edu*

**Abstract**

Denial-of-Service (DoS) and Distributed DoS (DDoS) attacks can cause serious problems in wireless networks due to limited network and host resources. Attacker traceback is a promising solution to take a proper countermeasure near the attack origins, to discourage attackers from launching attacks, and for forensics. However, attacker traceback in Mobile Ad-hoc Networks (MANETs) is a challenging problem due to the dynamic topology, and limited network resources. It is especially difficult to trace back attacker(s) when they are moving to avoid traceback. In this paper, we introduce the ATTENTION protocol framework, which pays special attention to MAC layer abnormal activity under attack. ATTENTION consists of three classes, namely, coarse-grained traceback, fine-grained traceback and spatio-temporal fusion architecture. For energy-efficient attacker searching in MANETs, we also utilize small-world model. Our simulation analysis shows 79% of success rate in DoS attacker traceback with coarse-grained attack signature. In addition, with fine-grained attack signature, it shows 97% of success rate in DoS attacker traceback and 83% of success rate in DDoS attacker traceback. We also show that ATTENTION has robustness against node collusion and mobility.

## 1. INTRODUCTION

The different types of denial of service attacks can be broadly classified into software exploits and flooding attacks. In software exploits, the attacker sends a few packets to exercise specific software bugs within the target's OS or application, disabling or harming the victim. On the other hand, in flooding attacks, one or more attackers send incessant streams of packets aimed at overwhelming link bandwidth or computing resources at the victim. In this paper, we mainly focus on flooding-type DoS/DDoS attack since it cannot be fixed with software debugging.

In flooding-type DoS/DDoS attack, attacker transmits a large number of packets towards victim with spoofed source address. For instance, in SYN Flood [2], at least 200-500 pps of SYN packets are transmitted to a single victim. UDP Echo-Chargen [4] and Smurf [3] also attacks victim using a large amount of packets with spoofed address. It is reported that DoS attack occurs more than 4,000 times per week and more than 600,000 pps of attack packets are used for attack in some cases [15]. In general, we can say the following characteristics of flooding-type DoS/DDoS attacks: (I) Traffic volume is abnormally increased during attack period. (II) Attackers routinely disguise their location using incorrect/spoofed addresses. (III) Such attacks may persist for tens of minutes and in some case for several days [1].

There are several IP traceback schemes proposed for the Internet such as packet marking [17][18], logging [16], ICMP traceback [7], etc [6]. Such traceback schemes developed for the fixed networks are not directly applicable to MANETs due to the following peculiar characteristics of MANETs: (1) In MANETs, there is no fixed infrastructure. Each node works as an autonomous terminal, acting as both host and a router. (2) Each node moves in and out, frequently changing network topology. (3) In general, network bandwidth and battery power are severely limited in MANETs compared to wired networks. (4) A node in MANETs has limited trust.

To perform efficient DoS/DDoS attacker traceback under such a harsh environment in MANETs, we specially pay attention to MAC layer abnormal activity under attack. Under flooding-type DoS/DDoS attack, *abnormal activities* (e.g., increased backoff time, increased number of frames, increased collision, etc) to access limited wireless medium are observed by *many overhearing nodes* around the attack path. We statistically characterize the MAC layer abnormality that is observed during DoS/DDoS attack and use the abnormality as attack signature for traceback. The attack signature is consistently observed on the attack path from attacker to victim, which enables us to track down attacker. The merits of MAC layer abnormality-based attacker traceback are multifold. First, we can track down attacker in spite of address spoofing using MAC abnormality-based attack signature. Second, the attack signature is observed by many neighbor nodes sharing the medium through overhearing. This overhearing can be efficiently used (i.e., majority voting) to prevent false/malicious reporting by compromised node or inside attacker. In addition, overhearing can be used for attacker traceback under node mobility. In MANETs, nodes frequently move in and out changing network topology. When a node that relayed attack traffic moves out, it is hard to trace back attack origin. In such case, we can use information from the nodes that overhear and stay in the region around attack path for traceback.

Another contribution of this paper is the analysis and traceback of mobile attackers. One of the main challenges in attacker traceback in MANETs is the mobility of nodes. Existing IP traceback schemes cannot be directly applied to MANETs under the presence of node mobility. When the attacker moves, this may confuse the victim between DDoS attack and mobile attack, which makes traceback fail. To track down mobile attacker, we introduce spatio-temporal fusion architecture. In the spatio-temporal fusion architecture, the relative location of attacker is estimated using multi-dimensional information, which is spatial and temporal information of attack signature. In addition, DDoS attack and mobile attack are effectively distinguished using spatio-temporal fusion architecture.

We summarize the contributions of this paper as follows.

- We propose a novel MAC layer abnormality detection mechanism that is effectively used as attack signatures with statistical characterization. We provide mechanisms robust to address spoofing, intermediate node collusion and false reporting. Further, we introduce the spatio-temporal fusion architecture; the first for mobile attacker detection.
- We provide two classes of traceback schemes (i.e., coarse-grained and fine grained) that are effective in both DoS and DDoS attacker traceback with low communication/processing overhead.
- We performe extensive simulation, with a rich set of scenarios, to validate the effectiveness of proposed protocol framework. Our results show a very high detection capability for DoS, DDoS and mobility cases.

The paper is organized as follows: In section 2, we briefly describe related IP traceback schemes. In section 3, we provide architecture overview of our proposal. In section 4 and 5 we provide two classes of attacker traceback schemes: coarse-grained traceback and fine-grained traceback. In section 6, we provide mobile attacker traceback architecture, which is spatio-temporal fusion architecture. Then, we present simulation results in section 7 and discussion and conclusion are provided in section 8 and section 9, respectively.

## 2. RELATED WORK

The method in [8] using controlled flooding tests network links between routers to determine the origin of the attack traffic. Downstream node intentionally sends a burst of network traffic to the upstream network segments. At the same time, it checks incoming attack traffic for any changes. From the changes and frequency of the incoming attack traffic, the victim can determine which upstream router the attack traffic is coming from. The same process is continued a level higher until finally reaching the attacker. Since this is a reactive method, the trace needs to be completed before the attack is over.

Packet Marking [17][18], and ICMP Traceback Message (iTrace) [7] attempt to distribute the burden of storing state and performing computation for IP traceback at the end hosts rather than in the network. For instance, in ICMP-based notification, router generates ICMP message containing information regarding where each packet came from and where it was sent. Then, routers notify the packet destination of their presence on the route. Collection of these messages can be used to trace the attack origin. ICMP traceback message uses ICMP but limits to generating ICMP message for every 20,000 packets (recommended). In Probabilistic Packet Marking (PPM), routers insert traceback data into each packet probabilistically so the number of packets that are marked at each router is enough for the reconstruction of attack path at the victim.

Logging scheme requires the routers to log meta-data in case an incoming packet proves to be offensive. Audited packet flow is logged at various points throughout the network and then used for appropriate extraction techniques to discover the packet's path through the network. To reduce the size of packet log and provide confidentiality, hash-based logging is proposed [16].

The existing schemes developed for the Internet are not directly applicable to MANETs due to the following reasons: (I) Intermediate relay nodes in MANETs can move in/out and may fail due to power outage, frequently changing network topology. In addition, each node in MANETs has limited trust due to the autonomous nature of nodes. Hence, traceback schemes that purely rely on relay nodes are problematic in terms of robustness and trust. (II) Storage capacity of each node is limited in MANETs. In packet marking and logging, large amount of per-packet information needs to be stored at either end-host or inside the network. (III) Existing schemes incur high processing load for attack path reconstruction. For instance, in iTrace, end host first searches the database, which stores path information of packets. Then, based on the per-packet information, end-host should run reconstruction algorithm to find out attack path. On the other hand, controlled flooding consumes a lot of bandwidth for traceback, which is highly undesirable in bandwidth-constrained MANETs.

SWAT [12,13] is the first traceback protocol developed for MANETs. SWAT consists of two main building blocks: Traffic pattern/volume matching and small world construction. It uses Traffic Pattern Matching (TPM) and Traffic Volume Matching (TVM) techniques to deal with address spoofing problem and utilize small-world model for efficient search. However, SWAT has the following drawbacks: (1) It cannot successfully trace back attacker when there exists high volume of background traffic. (2) SWAT fails to track down distributed DoS attackers. (3) SWAT also shows weakness under node collusion and false reporting since it relays only on relay nodes of attack traffic for traceback. (4) SWAT does not handle node mobility problems.

## 3. ARCHITECTURE OVERVIEW

When DoS/DDoS attack is detected by intrusion detection system, a victim initiates traceback process. The first step of traceback process is to characterize attack signature. The attack signature is characterized by two different classes: coarse-grained signature and fine-grained signature. In coarse-grained signature, attack is characterized by *regional* abnormality of MAC layer activity (increased busy time, increased frames, or increased collision, etc.). In fine-grained signature, attack signature is characterized by link-level abnormality using frame content information. There exists tradeoff between coarse-grained traceback (ATTENTION-CT) and fine-grained traceback (ATTENTION-FT). In coarse-grained traceback, we ignore content-level details. The fundamental assumption in coarse-grained traceback is that during DoS/DDoS attacks the regional wireless medium around attack path is heavily and abnormally loaded. This is generally true in bandwidth-constrained wireless networks. The abnormality is statistically characterized as attack signature and used for traceback. The advantage of coarse-grained traceback is that it is simple and computationally light to analyze attack traffic and perform traceback. The disadvantage is that it does not work in case the regional wireless medium is not abnormally loaded. Especially, in DDoS case, regional abnormality is low near the attack origins since only small amount of attack traffic is generated by each distributed attacker. In fine-grained traceback, content information such as previous hop MAC address, and next hop MAC address is used for traceback to filter out background traffic that does not contribute attack traffic. The obvious advantage is that we can trace back attacker even if there is low regional abnormality. On the other hand, the disadvantage is that it requires more computation overhead and memory for traceback.

Once the attack signature is characterized by the framework, victim node initiates efficient search process. By finding nodes in the neighbor, which observe similar or same attack signature, we can find the nodes that relayed the attack traffic. The process is continued recursively from the neighbor nodes of victim back to the attack origin. For efficient attacker searching, we use small world model. Helmy [10][11] found that path length in wireless networks is drastically reduced by adding a few random links (resembling a small world). These random links need not be totally random, but in fact may be confined to small fraction of the network diameter, thus reducing the overhead of creating such network. The random links can be established using contacts [11]. We effectively extend the contact architecture adding security functionality. Contact nodes are a set of nodes outside the vicinity, which are used as short-cut to build small world and provide wide view on entire network to the victim. Our contact selection scheme follows [11]. However, the functionality of contacts in ATTENTION are distinct, which will be articulated in Sec.4, 5, and 6.

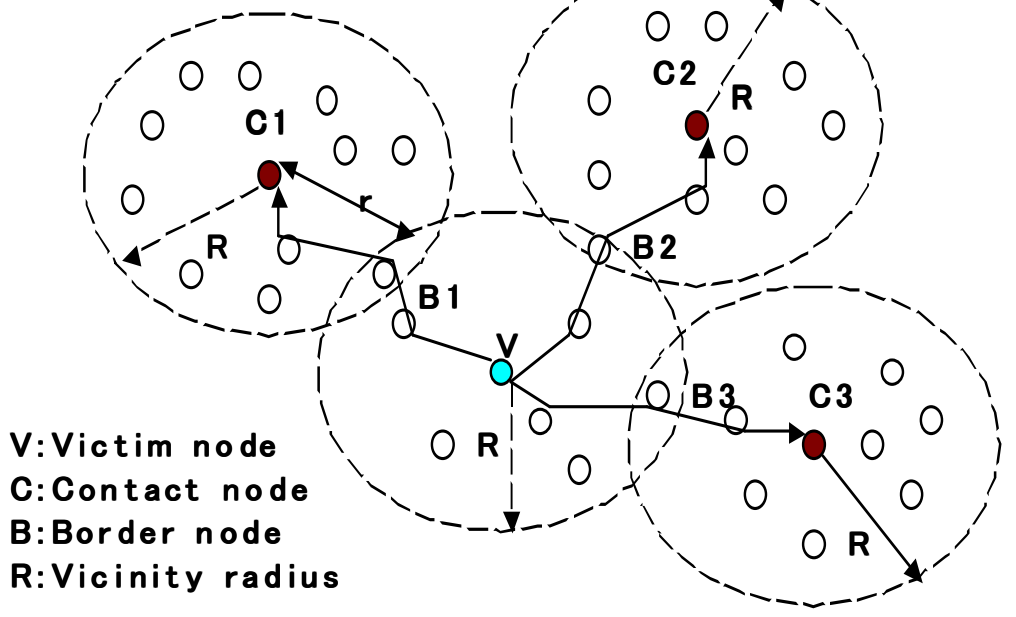

**[Figure 1] Each node has vicinity of radius *R* hops. A victim sends query with attack signature to its vicinity nodes and border nodes $B_i$. Then, the border nodes choose one of its borders $C_i$, to be the contact and sends query with attack signature.**

As shown in Fig.1, victim node, *V*, sends queries with attack signature to its vicinity nodes (nodes within radius *R*) and contacts (*C1, C2*, and *C3*). To send to the contacts, the victim node chooses three borders, *B1, B2* and *B3*, to which it sends the queries. The borders in turn choose three contacts at *r* hops away to which the borders forward the queries. If there is no node that observed (relayed or overheard) attack signature, it suppresses query. Otherwise, it sends next level query to the contact of contact. In doing so, we can perform directional search for DoS attacker traceback and multi-directional search for DDoS attacker traceback, where the search process has directionality towards attacker(s). Directional and multi-directional search significantly reduces communication overhead. We will verify the reduction in the simulation section.

To track down mobile attackers, we introduce spatio-temporal fusion architecture. The spatio-temporal fusion architecture consists of information gathering, and information fusion processes. These functionalities are implemented in contact architecture. Contact node gathers spatial/temporal attack information from its vicinity nodes. Then the information is correlated and classified in information fusion process. That is, movement of attacker is detected by correlating spatial/temporal continuity of attack signature movement. Fig. 2 summarizes the spatio-temporal fusion architecture and the detailed mechanism is explained in section 6.

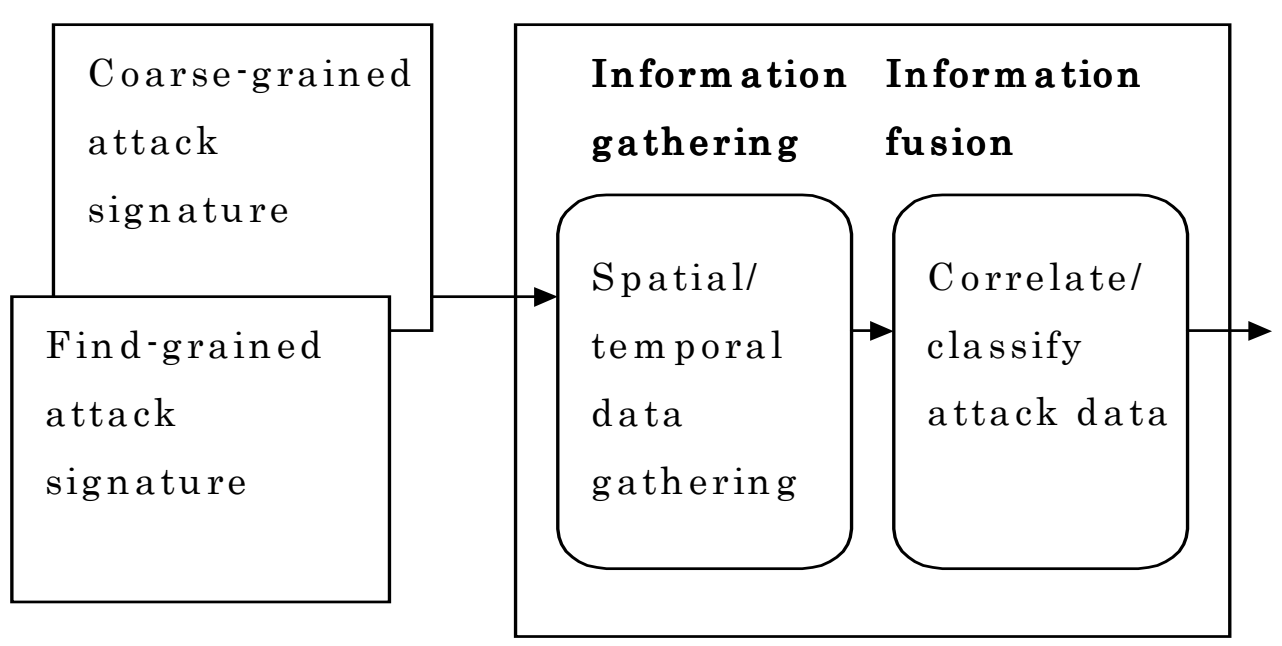

**[Figure 2] Building blocks of spatio-temporal fusion architecture for mobile attacker traceback**

## 4. COARSE-GRAINED TRACEBACK

Coarse-grained traceback consists of two phases. (I) In the first phase, we statistically characterize MAC layer abnormality (e.g., increased backoff time, increased number of Data/ACK/RTS frames, increased collision, etc), which is observed during attack. We define the abnormality as attack signature. The attack signature is consistently observed along the attack path from attacker to victim. (II) The second phase is an efficient search of intermediate nodes and attack origin using the attack signature.

### 4.1 MAC Abnormality Characterization Process

Once flooding-type DoS/DDoS attack is launched, a large volume of traffic is generated towards a victim to disable or harm network bandwidth or host resources. Consequently, we can observe abnormal MAC layer activity to capture medium and send large volume of packets. The abnormality is observed by neighbor nodes overhearing around the attack path. In this paper, we use 802.11 MAC mechanism, which is widely used for layer of MANETs and classify MAC layer activity into several components. Note that our scheme can be generally applied to other MAC protocols. Based on the classification, we analyze how the attack traffic causes abnormality in each component.

In the 802.11 protocol, the fundamental mechanism to access a medium is called Distributed Coordination Function (DCF). The DCF standard combines Carrier Sense Multiple Access/Collision Avoidance (CSMA/CA) with Request to Send/Clear to Send (RTS/CTS) handshake to avoid collisions. It works as follows: A node can only transmit when no carrier of a transmitting node is sensed in the vicinity. Otherwise, it has to defer its own transmission until the channel is determined to be idle. The node then requests channel by sending an RTS to the receiver, which, in turn, replies with a CTS. The nodes in the vicinity overhearing the RTS or CTS defer their own transmission for a period that is long enough for the subsequent DATA/ACK exchange. When the RTS/CTS handshake is completed, the sender commences data transmission. The receiver acknowledges the data with an ACK. If no CTS or ACK is received, the sender exponentially backs off, and retransmits the RTS or DATA. Under DoS/DDoS attack, we can observe several abnormalities.

■ *Increased collisions*

Increased collision can be inferred by several symptoms. (I) Increased retry count due to lack of ACK or CTS: frame or fragment has a single retry counter associated with it. Frames that are shorter than the RTS threshold have short retry count. Frames that are longer than the threshold are considered long frames and have long retry count. Frame retry counts begin at 0 and are incremented when a frame transmission fails. (II) Large contention window (*CW*). After each unsuccessful transmission, *CW* is doubled up to a maximum value $CW_{max} = 2^m * CW_{min}$, where $m$ is the number of attempt. (III) Long lifetime: when the first fragment is transmitted, the lifetime counter is started. When the lifetime limit is reached, the frame is discarded and no attempt is made to transmit any remaining fragments.

■ *Increased busy time*

A node monitors channel to check whether it is idle or not. If it is busy less than certain time interval, it cannot go into backoff stage and should defer. Frequent busy time and consequent transition from backoff state to defer stage are considered as symptom of heavy traffic.

■ *Increased frames*

As attack packets are increased, the number of corresponding data frames and ACK are increased. In addition, to access channel, the number of RTS and CTS frames are also increased.

We performed simulation to verify the abnormal behavior of MAC layer under DoS attack. We used *ns-2* for simulation with 50 nodes. The network size is 670m X 670m and DSDV is used for underlying routing protocol. Average distance between attacker and victim is 4 hops. In Fig.3, we varied the number of nodes that generate background traffic from 1 to 25 and measured the increase rate. Increase rate is defined as the ratio between abnormal behavior and normal behavior. For instance, when collision count under attack is $\eta$ and collision count under normal background traffic is $\eta'$, the increase rate is calculated as $\eta/\eta'$. In the simulation, attack traffic is generated as ten times of normal traffic size. As shown in the Fig.3, frame count and busy time show high increase rate when background traffic is low, and the increase rate decrease as background traffic increases. It is because as background traffic increases the attack traffic does not show drastic abnormality. On the other hand, in collision, increase rate is low when background traffic is low. It is because when there exists only attack traffic, collision rarely occurs. The increase rate of collision gradually goes up as background traffic increases and decreases after certain point. As we can see in Fig.3, there is clear MAC abnormality when background traffic is reasonable range. The abnormality is used for attacker traceback in ATTENTION.

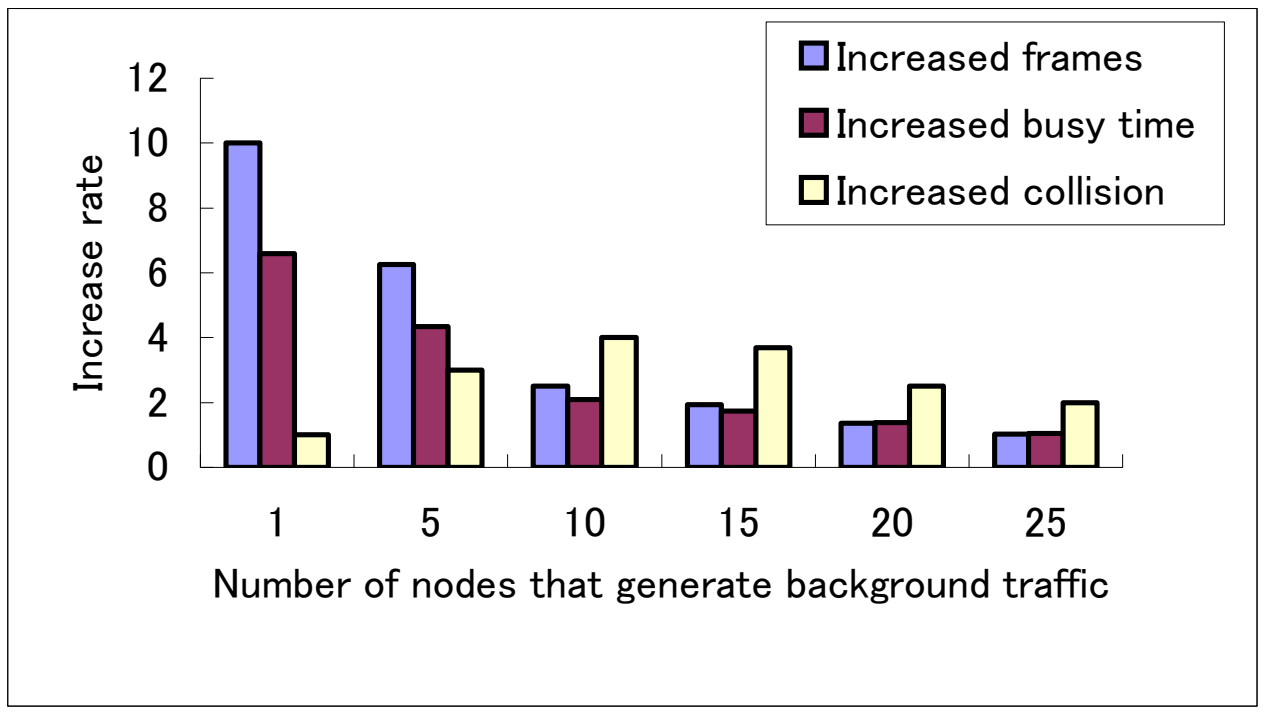

**[Figure 3] Abnormal increase of MAC activity component**

We first define the abnormality of MAC activity as the value that is outside the normal interval. Each node samples the MAC activity information (e.g., increased collision, busy time, frames, etc.) as time series data, $(x_1, x_2, \ldots x_n)$. Sampling number is chosen large enough to reflect the average network usage pattern and unit time for abnormality monitoring is chosen as 10 seconds in our scheme. After reaching maximum sampling number, the sampling process wraps to zero and start again. Based on the collected samples and using pivotal method [19], we calculate the normal interval with the confidence interval $100(1-\alpha)\%$ as follows.

$$\bar{x}_n \pm z_{\alpha/2}\left(\frac{\sigma}{\sqrt{n}}\right) \approx \bar{x}_n \pm z_{\alpha/2}\left(\frac{s_n}{\sqrt{n}}\right) \quad \text{(Eq.1)}$$

Where, $\bar{x}_n$ is the sample mean and σ is the standard deviation. Since the value of σ is unknown, the sample standard deviation $s_n$ is used.

The mean and variance is calculated and updated as follows: The average $\bar{x}_n$ of the time series data given $n$ points is obtained as follows

$$\bar{x}_n = \frac{1}{n}\sum_{1}^{n} x_i \text{ (Eq.2)}$$

If a new point $x_{n+1}$ is measured and it is within normal range, we can recompute $\bar{x}_n$ , but it is more efficient to use the old value of $\bar{x}_n$ and make a small correction using $x_{n+1}$. The correction is easy to derive, since,

$$\bar{x}_{n+1} = \frac{1}{n+1}\sum_{1}^{n+1} x_i = \frac{n}{n+1}\left(\frac{1}{n}\sum_{1}^{n} x_i + \frac{1}{n} \quad x_{n+1}\right) \text{ (Eq.3)}$$

And so, $\bar{x}_{n+1}$ can be written as

$$\bar{x}_{n+1} = \frac{n}{n+1}\bar{x}_n + \frac{1}{n+1}x_{n+1} = \bar{x}_n + K(x_{n+1} - \bar{x}_n) \text{ (Eq.4)}$$

Where, $K=1/n+1$ as gain factor. The gain $K$ adjusts how big the correction will be. We can also recalculate recursively the quadratic standard deviation (variance) of the time series data. Given $n$ points, the quadratic standard deviation is given by:

$$s_n^2 = \frac{1}{n}\sum_{1}^{n}(x_i - \bar{x}_n)^2 \text{ (Eq.5)}$$

If a new point $x_{n+1}$ is measured, the new variance is

$$s_{n+1}^2 = \frac{1}{n+1}\sum_{1}^{n+1}(x_i - \bar{x}_{n+1})^2 = \frac{1}{n+1}\sum_{1}^{n+1}(x_i - \bar{x}_n - K(x_{n+1} - \bar{x}_n)^2) \text{(Eq.6)}$$

The whole expression reduces to

$$s_{n+1}^2 = \frac{n}{n+1}(s_n^2 + K(x_{n+1} - \bar{x}_n)^2) = (1-K)(s_n^2 + K(x_{n+1} - \bar{x}_n)^2) \text{ (Eq.7)}$$

The whole process can now be cast into as a series of steps to be followed iteratively. Given the first $n$ points, and our calculation of $\bar{x}_n$ and $s_n$ , then

(1) When a new point $x_{n+1}$ is measured, we compute the gain factor $K=1/(n+1)$.
(2) We compute the new estimation of the average

$$\bar{x}_{n+1} = \bar{x}_n + K(x_{n+1} - \bar{x}_n) \text{ (Eq.8)}$$

(3) We compute also a provisional estimate of the new standard deviation

$$s_n'^2 = s_n^2 + K(x_{n+1} - \bar{x}_n)^2 \text{ (Eq.9)}$$

(4) Then, we find the correct $s_{n+1}$ using the correction

$$s_{n+1}^2 = (1-K)s_n'^2 \text{ (Eq.10)}$$

(5) Finally, we calculate the normal range

$$(\bar{x}_n - z_{\alpha/2}(\frac{s_n}{\sqrt{n}}), \bar{x}_n + z_{\alpha/2}(\frac{s_n}{\sqrt{n}})) \text{ (Eq.11)}$$

The iterative computation is used for updating average, standard deviation and consequent normal range. When the new value is outside the normal range, we define it as abnormality. The time series data, $(x_1, x_2, \ldots x_n)$ which is outside the normal range is defined as *attack signature* and used for traceback. Each node monitors MAC layer activity to detect the abnormality. Once the abnormality is observed, a node characterize the abnormality as attack signature and it is logged for traceback. The abnormal range is excluded from calculating normal range.

To find optimal indication factor for attack signature, we use statistical test (categorical data analysis) [19]. The optimal indicator should possess the following property: "*Majority* of neighbor nodes around the attack path observe *abnormal MAC activity*". We use two-way contingency table to evaluate the dependency of MAC layer activity on attack traffic. In two-way contingency table, data is classified according to the direction (row and column) of classification, according to two qualitative variables. In our test, row is the normal/abnormal and column is attack region/non-attack region as in table 1. Attack region is the area around the attack path where nodes can overhear attack traffic activity. In the contents of table, the number of nodes that observes corresponding MAC layer abnormality is written. $N$ is the total number of nodes in the networks.

**[Table 1] Two-way contingency table**

| | Non-attack region | Attack region | Totals |
|---|---|---|---|
| Normal | $n_{11}$ | $n_{12}$ | $n_{1.}$ |
| Abnormal | $n_{21}$ | $n_{22}$ | $n_{2.}$ |
| Totals | $n_{.1}$ | $n_{.2}$ | $N$ |

Then, we set the following null and alternative hypothesis to test the dependency of two classifications.

$H_0$: The two classifications are independent
$H_a$: The two classifications are dependent

Test statistic: $$\chi^2 = \sum_{j=1}^{c}\sum_{i=1}^{r}\frac{[n_{ij} - \hat{E}(n_{ij})]^2}{\hat{E}(n_{ij})} \quad \text{(Eq.12)}$$

Where, $\hat{E}(n_{ij}) = \frac{n_{i.}n_{.j}}{n}$, $n_{i}.$ is total for row $i$ and $n_{.j}$ is total for column $j$. The rejection region where we can conclude that two classifications are dependent is as follow.

$$\chi^2 > \chi_\alpha^2$$

Where $\chi^2$ is chi-square probability distribution with $(r-1)(c-1)$ degree of freedom and $\alpha$ is the probability of a type I error (a type I error is made if $H_0$ is rejected when $H_0$ is true). Intuitively, $\chi^2$ shows high value as the percentage of nodes that observes abnormality increases.

In Fig.4, we show the $\chi^2$ value of each abnormality component. The threshold $\chi_\alpha^2$ is 5.02 with 97.5% confidence interval. We can infer that if the $\chi^2$ value is above the threshold, there exists dependency (reject the null hypothesis $H_0$). As shown in the Fig.4, when the number of nodes that generates background traffic is low (less than 10% of entire network nodes), the dependency of frame count and busy time on attack is high. As background traffic increases the dependency is decreased since there is little difference between attack traffic and background traffic. On the other hand, the dependency of collision on attack is low when background traffic is low since there is only small background traffic that collides with attack traffic. When the background traffic is high (above 30), the dependency is low. It is because there is only little difference between attack traffic and background traffic. In other region, we can constantly observe high $\chi^2$ value ($\chi^2 > \chi_\alpha^2$), which means that attack traffic have clear impact (dependency) on the overhearing nodes around attack route.

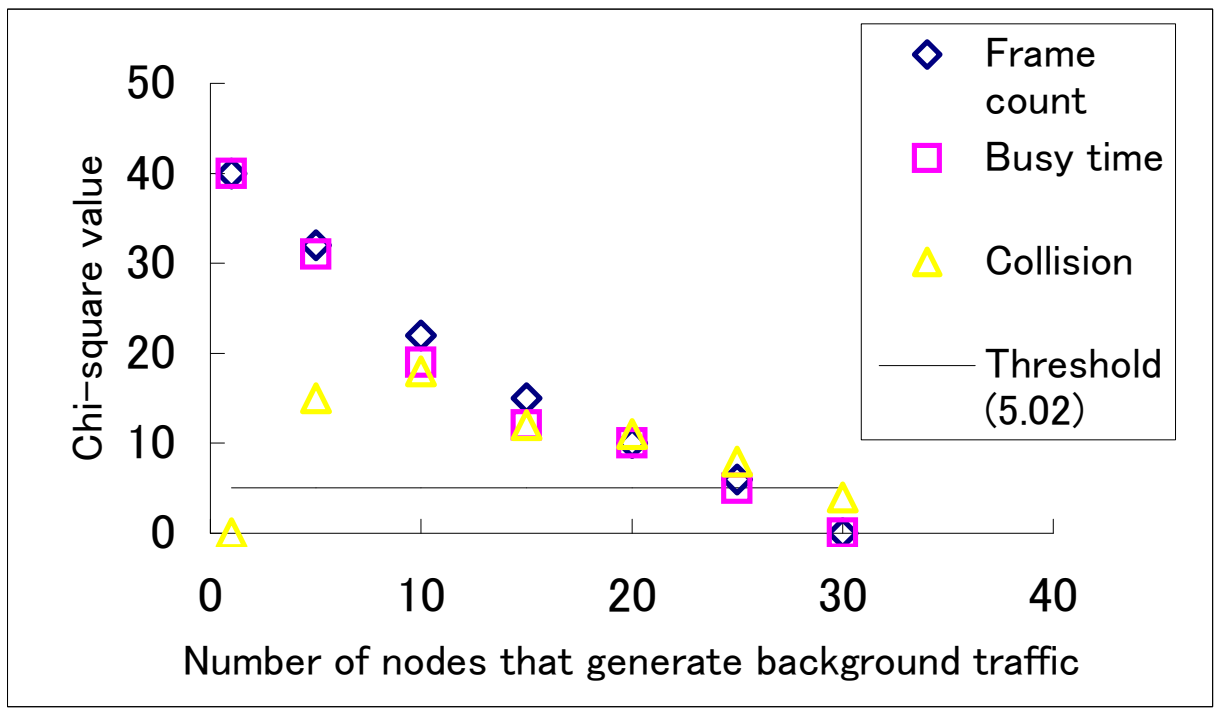

**[Figure 4] Dependency of MAC layer activity on attack**

## 4.2 Abnormality Matching using KS test

We are interested in using the Kolmogorov-Smirnov (KS) statistic $D_n$ [18] to test the hypothesis that the two abnormality, $F_n(x)$, $F_0(x)$ is matching. $F_0(x)$ corresponds to reference abnormality, which is included in query message, and $F_n(x)$ is the candidate abnormality observed by vicinity nodes.

$$D_n = \sup_x [\ |F_n(x) - F_0(x)|\ ] \qquad \text{(Eq.13)}$$

$$H_0 : F_n(x) = F_0(x)$$
$$H_a : F_n(x) \neq F_0(x)$$

We accept $H_0$ if the distribution function $F_n(x)$ is sufficiently close to $F_0(x)$, that is, if the value of $D_n$ is sufficiently small. The hypothesis $H_0$ is rejected if the observed value of $D_n$ is greater than the selected critical value that depends on the desired significance level and sample size. When the $H_0$ is accepted (sufficiently similar), we can infer that the abnormality is matching, meaning that the attack traffic is traversed the two contact region. In our scheme, when there are multiple matching results, we select the one, which shows the smallest $D_n$ value as the attack traffic relay region. This process is performed recursively from level-1 contacts towards attack origin.

## 4.3 Attacker Searching Process

We describe the DoS attack traceback scheme as follows: (1) when a victim node, *V*, detects attack such as SYN flooding, it first extracts attack signature described by MAC layer abnormality. It then sends a query to the nodes within its vicinity and level-*1* contacts specifying the depth of search (*D*) large enough to detect an attacker. The query contains sequence number (*SN*) and attack signature. (2) As the query is forwarded, each node traversed records the *SN*, and *V*. If a node receives a request with the same *SN* and *V*, it drops the query. This provides for loop prevention and avoidance of re-visits to the covered parts of the network. (3) In case KS test is passed, meaning that there exist vicinity nodes of contacts that observe similar attack signature, the first step of trace is completed. For instance, victim (*V*) sends query to the vicinity nodes and 2 level-*1* Contacts (*CL_1a* and *CL_1b*) around the victim in Fig. 5 (transmission arrows to vicinity nodes by each contact are omitted in the figure). Then, one level-*1* (*CL_1b*) contact reports to the victim that some of its vicinity nodes observed low $D_n$ value in KS test. (4) Next, only the contact, *CL_1b*, that observes traffic signature matching in its vicinity sends next level query to level-*2* contacts (*CL_2c*, and *CL_2d*) with the partial attack path appended to the query. It also reduces *D* by 1. This processing by contact is called *in-network processing*. Other contacts that do not have relay nodes of attack traffic in their vicinities, suppress forwarding the query *(query suppression)*. This results in *directional search* towards the attacker. (5) When there is no more contact report or no other nodes outside the vicinity, the last contact (*CL_2c*) reports the complete attack route to the victim.

Our scheme is based on majority voting. That is, even if some nodes move out from the attack route or are compromised by attackers, we can still find an attack route using available information from good nodes staying in the vicinity. This majority voting becomes possible in ATTENTION since MAC layer abnormality is observed by many overhearing nodes around attack path.

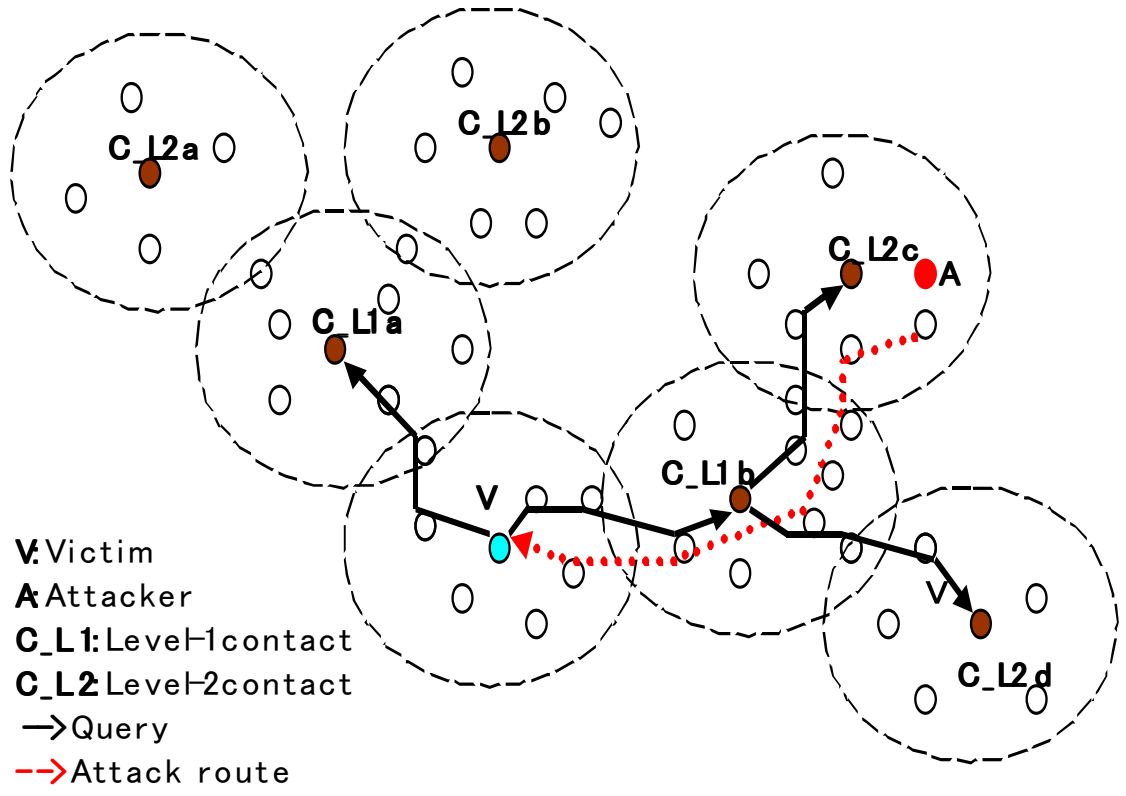

**[Figure 5] Victim (*V*) sends queries with attack signature to the first level contacts, (*CL_1a, CL_1b*). Only *CL_1b* that observed matching traffic signature within vicinity sends next level queries to level-2 contacts (*CL_2c, CL_2d*). *CL_1a* suppresses further query. *CL_2c* sends final attack route to the victim.**

## 5. FINE-GRAINED TRACEBACK

There are several drawbacks of coarse-grained traceback scheme. First, if the attack traffic does not show regional abnormality, it is difficult to trace back attacker using coarse-grained abnormality matching. Especially in DDoS attack, attacker orchestrates attack so that each distributed node contributes only small portion of attack traffic to the victim. Second, if there exists a large amount of background traffic, attack traffic shows low matching level (high $D_n$ in KS test) along the attack path. The fundamental problem of coarse-grained traceback is that simple count information of *regional* MAC layer abnormality includes a lot of *noise* factors by background traffic. It is necessary to capture fine-grained traffic information to filter out the noise factors for more accurate traceback. We introduce fine-grained traceback scheme, which utilizes content information inside frame. Fine-grained traceback consists of two phases: (1) *Link-level abnormality detection* (2) *Directional noise reduction*

Each node monitors link-level abnormality instead of region-level abnormality. That is, a node monitors abnormality with the tuple *(Src_addr, Dest_addr, Φ),* where *Src_addr* is the MAC address of sender and *Dest_addr* is the MAC address of destination, which is observed by relay/overhearing nodes. Note that MAC layer information is not susceptible to network layer spoofing (e.g., source IP address) by attacker. We assume that intermediate nodes, which only relay traffic, do not spoof its own MAC addresses. $\Phi$ is the time series of number of frames (i.e., $(N_1,N_2,...,N_t)$, $N_i$=the number of frames at time slot *i*) from *Src_addr* to *Dest_addr* during last time frame [0,*t*]. If one of links shows abnormality, which is outside the normal range (Eq.11), it logs the attack signature. Link-level abnormality has the following advantages: (I) It captures the abnormality more sensitively compared to region-level monitoring since link-level traffic includes smaller background traffic. (II) The matching test shows more accurate result since noise factors caused by background traffic are reduced.

In addition, fine-grained signature filters out directional noise, through relay continuity discovery. Relay continuity discovery is the process to find the list of continuous nodes that relayed attack traffic. Let $(Src_addr_i, Dest_addr_i, \Phi_i)$ be the observation by node *i* and $(Src_addr_j, Dest_addr_j, \Phi_j)$ be the observation by node *j*. When $Src_addr_j = Dest_addr_i$ and $\xi_i \approx \xi_i$, we can infer that attack traffic is relayed from node *i* to node *j*. The nodes that do not have continuity which ultimately end up at the victim are excluded in the traceback. Note that the continuity discovery process only checks local continuity. Hence, it does not completely exclude traffic that is not ultimately heading to the victim. However, we can reduce much noise by eliminating the links that do not have relay continuity.

Under DDoS attack, attack traffic is merged in various parts of networks or at the victim. Hence, partial attack traffic should be detected for distributed attacker traceback. To find the partial attack traffic, we perform the following combinational matching test: Let $\Phi_i$, $\Phi_j$, and $\Phi_k$ be the time-series data, which shows abnormality and has continuity to the victim. We call the time-series data, which shows abnormality, as candidate attack signature. Then, combinational abnormality matching test is performed between the time-wise summation of each combination of candidate attack signatures and attack signature ξ. There can be *S* number of combinations from *K* (3 in the example) candidate attack signatures as follows.

$$S = \sum_{i=1}^{K} {}_K C_i \text{ (Eq.14)}$$

We find the distributed attack route by finding high matching level between each combination and attack signature (e.g., $\Phi_i+\Phi_j \approx \xi$). Since attack signature is formed by the partial attack traffic, background traffic is to be excluded through the combinational matching test.

Overall fine-grained attacker traceback scheme to track down DDoS attackers is explained as follows. (1) First, a victim node identifies attack signature. Unlike DoS attack, attack signature of DDoS attack consists of multiple signatures that come from different neighbors of the victim. That is, the victim, identifies multiple attack signatures ($\Phi_1, \Phi_2,...,\Phi_n$), *which* come from different sources (i.e., $Src_addr_1, Src_addr_2,..., Src_addr_n$) during attack period *t.* (2) Once the attack signatures are identified, search process begins with relay continuity discovery. That is the victim sends attack signature matching query to its vicinity nodes and level-1 contacts. Candidate find-grained attack signatures that have continuity are returned to the victim. (3) Then victim performs attack signature matching test between the links that has continuity. Note that, unlike DoS attacker traceback, matching test should be done at the victim node. (4) If there exist multiple sender nodes that are sending abnormal traffic to the same destination, we can infer that attack traffic is merging and matching test should done between each combinations of candidate attack signature and attack signature. (5) The combination of candidate attack signature that shows the highest matching level becomes new branch attack signature. (6) The process is recursively repeated towards attack origins

Note that DoS attacker can also be traced back with fine-grained signature, which may increase traceback success rate. However, it incurs increased processing load and memory overhead.

## 6. MOBILE ATTACKER TRACEBACK

In MANETs, nodes move changing network topology. DoS/DDoS attack under such a mobile scenario causes several problems/illusions in traceback. (I) When attacker changes its location and performs DoS attack, victim might confuse mobile attack as DDoS attack, which leads to false traceback result. To distinguish DDoS attack and mobile attack, we take *age information* into consideration. Age is defined as the *first* and *last* (or *most recent*) time $(t_S, t_L)$ the abnormality is observed. In addition, we introduce *spatial continuity* and *temporal continuity* detection for mobile attacker traceback. (II) We need to estimate the current location of attacker to take proper countermeasure near the attack origin and for forensics. We introduce *relative* attacker positioning mechanism, which does not require geographic information. We assume majority of intermediate nodes, which overhears attack signature stays in the vicinity of attack route. We call our mobile attacker traceback system as *spatio-temporal fusion architecture.* The basic step is explained as follows.

(1) Once attack is detected by intrusion detection system, victim initiates traceback process by sending attack signature queries to its contacts.
(2) Contact, in turn, sends queries to its vicinity nodes to find the nodes that observed abnormality. The nodes that detect the abnormality report the attack signature with age information to the contact.
(3) The contact aggregates the attack information from its vicinity nodes and reports it to the victim. If there exist next-level contacts, the contact sends the abnormality query to the next-level contact. The process is continued recursively.

(4) The victim uses the information from all the contacts and infers the mobility of attack. In addition, a victim estimates the current location of attacker based on the gathered information.

The spatio-temporal fusion architecture is divided into information gathering, and information fusion process.

## 6.1 Information Gathering

Similar to coarse-grained and fine-grained traceback, attack information plus *age information* is gathered by contact from vicinity nodes. More formally, spatio-temporal attack signature consists of $(\xi, t_S, t_L, S)$. $\xi$ is the attack signature that is either coarse-grained or fine-grained. $S$ is the relative position of attacker (e.g., 2 hops away from level-*1* contact $i$). This tuple $(\xi, t_S, t_L, S)$ is effectively used to classify attack type (e.g., DDoS attack, mobile attack, etc).

The contacts that observe attack signature expand the queries to next level contacts and other contacts suppress query, which results in (multi) directional information gathering. The difference between spatio-temporal architecture and coarse-grained/fine-grained traceback is that all the attack information from every level of contact is returned to the victim and the victim analyzes the information.

## 6.2 Information Fusion

Information fusion is the process to correlate and analyze the spatio-temporal information obtained through information gathering process. Before we describe information fusion process, we define two *atomic components* to detect mobile attack, namely *spatial relation* and *temporal relation*. Different types of attacks (e.g., mobile attack, DDoS attack, etc) can be effectively distinguished using the spatial/temporal relation. For instance, in mobile DoS attack, the attack signature ($\xi$) is observed in a spatially continuous manner (Fig.6 (a)). In general, spatial discontinuity is observed in DDoS attack as in Fig.6 (b). Note that DDoS attack can also show spatial continuity. In that case, we can distinguish DDoS attack and mobile DoS attack using temporal relation described later in this section. Note that *each cell logically corresponds to contact vicinity* in ATTENTION architecture.

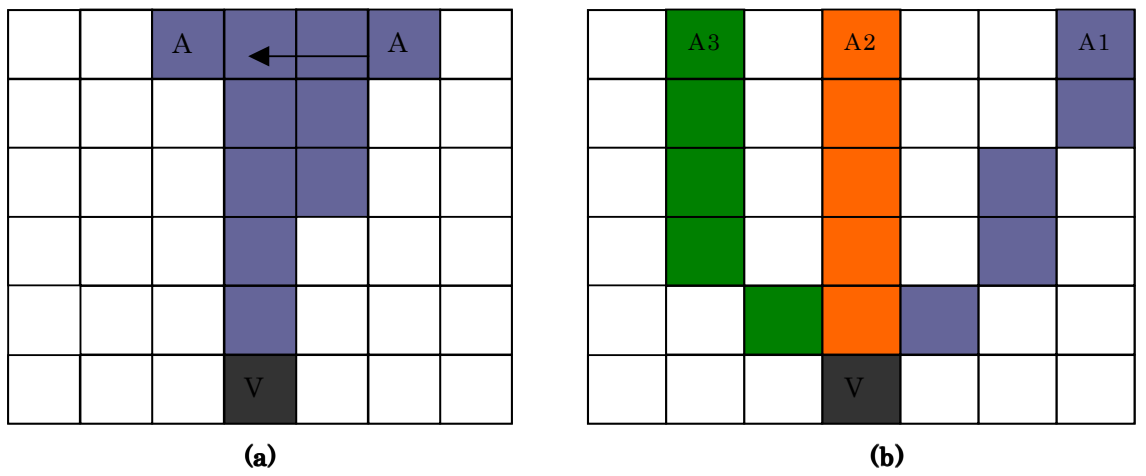

**[Figure 6] (a) Spatial continuity of mobile DoS attack. Attacker, *A*, is moving from right to left attacking victim *V*. (b) Spatial discontinuity of DDoS attack. Attackers, *A1*, *A2*, and *A3* are launching attacks towards victim *V*.**

To quantatively formulate spatial relation, we define Spatial Relation Factor (SRF) as follows.

$$SRF = \alpha \bullet \frac{1}{\dfrac{\sum_{i_x=1}^{N_x} \sum_{j_y=1}^{N_y} D_S(i_x, j_y, \xi)}{P}} = \frac{\alpha \bullet P}{\sum_{i_x=1}^{N_x} \sum_{j_y=1}^{N_y} D_S(i_x, j_y, \xi)} \quad \text{(Eq.15)}$$

Where,

$$\alpha = \frac{n_S}{N_x + N_y} \quad \text{(Eq.16)}$$

$N_x$ is the total number of vicinity nodes of contact $x$ and $N_y$ is the total number of vicinity nodes of contact $y$. $n_S$ is the number of nodes that observe similar attack signature, $\xi$, in the vicinity of contact $x$ and the vicinity of contact $y$. $i_x$ is a vicinity node of contact $x$ and $j_y$ is a vicinity node of contact $y$. $D_S(i_x, j_y, \xi)$ is the hop count between node $i_x$ and $j_y$ that observe the attack signature $\xi$. The hop count information is obtained using underlying routing table. $D_S(i_x, j_y, \xi)=0$ if node $i_x$ and $i_y$ do not observe the similar attack signature, $\xi$. $P$ is the total number of tuples where $D_S(i_x, j_y, \xi)>0$. By high α value, we can infer that attacker is occurring near the central region of $x$'s vicinity and $y$'s vicinity. It is because more neighbors can overhear the abnormality when attacker passes through near the central region of the contact's vicinity. When α is small, we can infer that the attacker is not passing through the central region of the contact's vicinity or the matching report is not reliable (false reporting). In addition, when attacker moves from vicinity of $x$ to vicinity of $y$, we can observe small $D_S(i_x, j_y, \xi)$. Consequently, when $x$ and $y$ is not adjacent contact and $i_x$ and $i_y$ is far away, large $D_S(i_x, i_y, \xi)$ is obtained, which leads to low SCF.

We classify temporal relation of attack into temporal continuity, temporal discontinuity, and temporal overlap as in Fig.7. *T0*, *T1*, and *T2* are the time slot at which the attack is observed. In Fig.7 (a), attack signature ($\xi$) is observed at *T0, T1*, and *T2* time slots continuously (Temporal continuity). On the other hand, temporal discontinuity is defined as in Fig.7 (b). The attack signature ($\xi$) is observed at discontinuous time slots *T0* and *T2.* Temporal overlap (Fig.7(c)) implies that attack is occurring simultaneously.

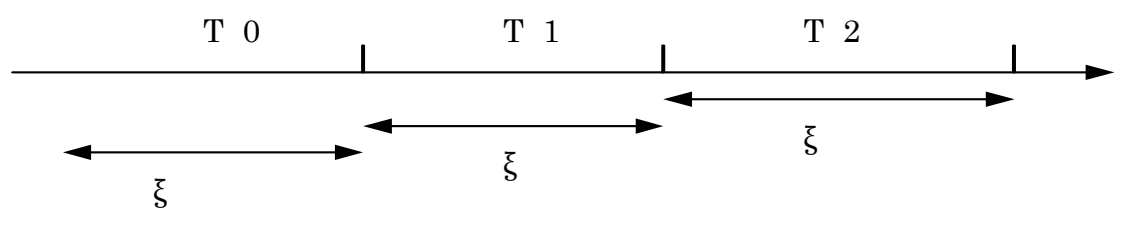

(a) Temporal continuity

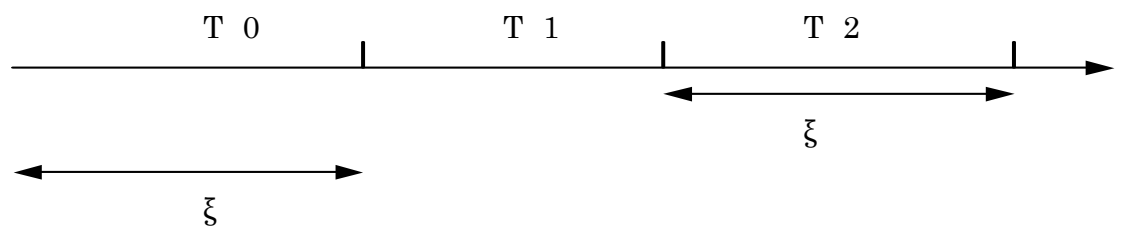

(b) Temporal discontinuity

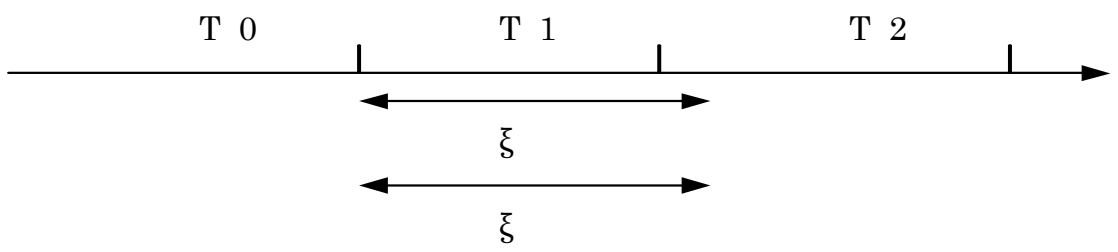

(c) Temporal overlap

**[Figure 7] Three types of temporal relation of attack**

We quantatively formulate the temporal relation as Temporal Relation Factor (TRF).

$$TRF = \alpha \bullet \frac{1}{\frac{\sum_{i_x=1}^{N_x}\sum_{j_y=1}^{N_y} D_T(t_L(i_x), t_S(j_y), \xi)}{P}} = \frac{\alpha \bullet P}{\sum_{i_x=1}^{N_x}\sum_{j_y=1}^{N_y} D_T(t_L(i_x), t_S(j_y), \xi)} \quad \text{(Eq.17)}$$

Where, $D_T(t_L(i_x), t_S(j_y), \xi))$ is the difference between the start time (i.e., $t_S(i_y)$) when attack signature is observed at node $i_y$ and the last (or most recent) time (i.e., $t_L(i_x)$) when the attack signature is observed at $i_x$ where $t_L(i_x)<t_S(i_y)$. Under mobile attack, temporal continuity is observed and TRF becomes large since $D_T(t_L(i_x), t_S(j_y), \xi))$ becomes small. On the other hand, under DDoS attack, temporal overlap is observed and TRF becomes small value since $D_T(t_L(i_x), t_S(j_y), \xi))$ becomes large.

We use SRF and TRF metrics to infer attack type as follows: (I) When high SRF and high TRF is observed, we can infer that mobile attack has occurred. (II) When high SRF, and low TRF are observed, we can infer that DDoS attack has occurred from geographically clustered attackers. (Clustered DDoS attack) (III) When high SRF and extremely low TRF (<ε) are observed, we can infer that intermittent attack has occurred from clustered attackers. (IV) When low SRF, and high TRF are observed, we can infer that attack has occurred from geographically spread attackers with temporal continuity. (V) When low SRF and low TRF are observed, we can infer that DDoS attack from geographically spread attackers has occurred (spread DDoS attack). (VI) When low SRF and extremely low TRF (<ε) are observed, we can infer that intermittent attack has occurred from geographically spread attackers. Appropriate threshold for SRF and TRF is examined in the simulation section.

In addition, the current Relative Location (RL) of attacker can be estimated with following equation.

$$RL = \frac{\sum_{i=1}^{N} D_S(i, c_i, t_L(i), \xi)}{N} \quad \text{(Eq.18)}$$

Where, $D_s(i,c_i,t_L(i), \xi)$ is the distance (hop count) between contact $c_i$ and its vicinity nodes, $i$, which observe attack signature at the largest (the most recent) time, max $(t_L(1),\ldots, t_L(N))$. $D_s(i,c_i,t_L(i), \xi)=0$ when $t_L(i) \neq$ max $(t_L(1),\ldots, t_L(N))$. $N$ is the number of nodes that observes the attack signature and $D_s(i,c_i,t_L(i), \xi)>0$. It estimates the relative location of attacker (i.e., number of hops from contact). The rational of the positioning is that the attacker is located approximately at the center of overhearing nodes. Note that *RL* has robustness against attacker's MAC address spoofing since we do not rely on MAC address inscribed in frame for the positioning. However, it does not provide directionality.

We provide examples and algorithm to track down mobile attack based on the above classification in the following.

■ Mobile DoS attack

Fig. 8(a) shows an example of mobile DoS attack detection using the TRF and SRF metrics. In the figure, attacker moved from region 10→9→8→7. Attack path from each cell is as follows: (10→6→4→2→1→v), (9→6→3→2→1→v), (8→5→3→2→1→v), (7→5→3→2 →1→v). A victim finds first level temporal/spatial relation of attack at region 3, and 4. In region 3 and region 4, high TRF and high SRF (4→3) are observed. At this point, we can infer that mobile attack is occurring. Similarly, in region 5 and 6, high TRF and high SRF are observed. Lastly, in region 7, 8, 9, 10, high TRF and SRF are observed, which leads us to conclude that attacker is moving and currently located in the region 7. Relative location of attacker is calculated using Eq.18 at region 7. Vertical or diagonal movement of attacker can be detected similarly.

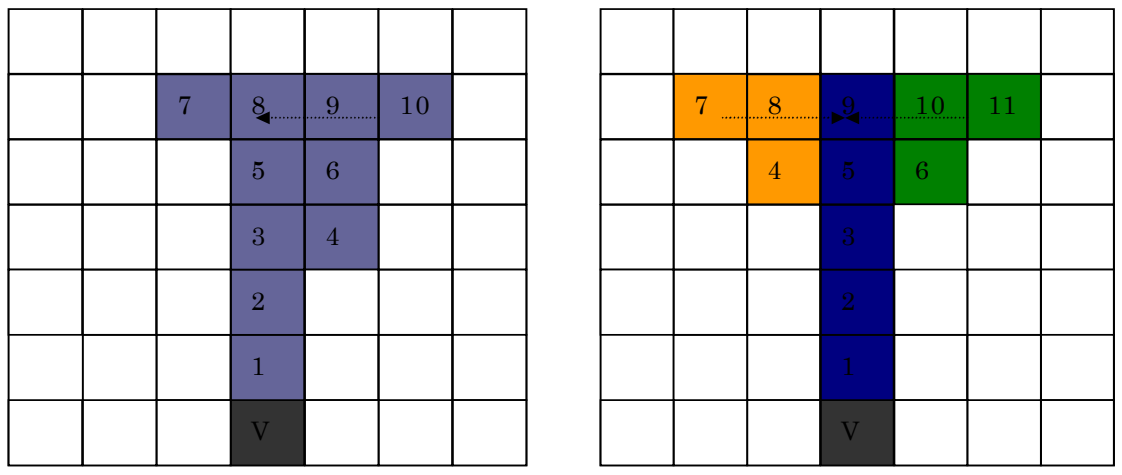

**(a) Mobile DoS attack** **(b) Crossing mobile DDoS attack**

**[Figure 8] Illustration of information fusion process for mobile DoS/DDoS attack detection**

■ Mobile DDoS attack

Basically, mobile DDoS attack can be detected and traced through separate path with the same mechanism as mobile DoS attacker traceback. The difficult problem in mobile DDoS attack occurs when two attacker are crossing each other as in Fig.8 (b). The crossing mobile DDoS attack can be detected by using TRF and SRF metrics plus the detection of attack signature surge. Attack signature surge is observed since two attack traffic are merged when crossing each other. For instance, in Fig.8 (b), first attacker is moving from 7→8→9 and the second attacker is moving from 11→10→9 and attack traffic is merged on the path from 9→6→3→2→1. Region 4, and 5 observe high SRF and high TRF with attack signature $\xi_1$. Region 5, and 6 also observe high SRF and high TRF with attack signature $\xi_2$. The relation enables us to infer mobile attack has occurred in region 4, 5, and 6. In addition region 5 observes the surge of attack traffic ($\approx\xi_1+\xi_2$), which enable us to infer the cross of mobile attack traffic. Similarly region 7,8 and 9 observe high SRF and high TRF. Region 11,10, and 9 observe high SRF and high TRF. Region 9 observes the surge of attack signature. Relative location of attacker is calculated using Eq.18 at contact region 9. The overall algorithm to detect and trace mobile DDoS attack is summarized in Fig.9.

**Procedure at victim *v***

STEP 1: Detect flooding-type DoS/DDoS attack.

STEP 2: Send attack signature query to level-*1* contacts.

STEP 3: If there are multiple signature reports from contacts, calculate SRF and TRF.

STEP 4: If SRF > SRF_thresh and TRF > TRF_thresh between contact $c_{1a}$ and $c_{1b}$, infer that attacker is moving from region $c_{1a}$ to region $c_{1a}$.

STEP 5: Check signature surging. If the surging exists, infer that multiple attackers are crossing.

STEP 6: Wait matching report from higher level contact and perform SRF and TRF calculation.

STEP 7: If there is no more report after receiving level-*N* contact report, infer the current relative position of attack (Eq.18) at level-*i* contact, which has the largest age.

**Procedure at intermediate contact $c_i$**

STEP 1: Receive attack signature query from contact $c_{i-1}$ or victim.

STEP 2: Gather abnormality information from its vicinity nodes.

STEP 3: If abnormality exists, report the attack signature to the victim.

STEP 4: If there exists contact of contact ($c_{i+1}$) and abnormality is observed, send next-level query $c_{i+1}$. Otherwise suppress query.

**[Figure 9] Overall algorithm to detect mobile attack**

## 7. SIMULATION RESULTS

We have performed extensive simulations using *ns-2* to evaluate the effectiveness of the proposed traceback schemes, specifically coarse-grained traceback, fine-grained traceback, and spatio-temporal fusion architecture, with various parameter spaces. Transmission range of each node is set 150m and networks size is 2750mX2750m. We repeated each simulation 10 times in random topology and calculated the average value. The evaluation metrics that we measured in the simulation are traceback success rate, false positive rate, communication overhead, SRF and TRF. We set *NoC* (Number of Contacts) = 6, *R* (vicinity radius) = 3, *r* (contact distance) = 3, *d* (search depth) =5 for contact selection. DSDV is used as underlying routing protocol. DoS attacker is 17 hops away from victim, and DDoS attacker is 10 hops away from victim. Background traffic is generated with the volume of 7.5% of attack traffic (i.e., if attack traffic=500pps, then, background traffic=(7.5*500pps)/100≈38pps) from random nodes.

### ■ Coarse-grained traceback

Fig.10 shows the success rate of DoS attacker traceback. *x* axis represents the percentage of nodes (out of 1,000 nodes in the entire network) that generate background traffic. In frame count and, busy time, traceback success rate is very high when background traffic low or medium. However, as background traffic is increased the success rate is decreased due to the following reasons: (I) Abnormality matching level is decreased since more noise is included by background traffic. (II) Attack traffic does not show abnormal characteristics. Hence, nodes around the attack route cannot capture the abnormality. On the other hand, in collision, the traceback success rate is very low when background traffic is low since collision seldom occurs. However, as background traffic is increased the success rate is also increased. Then, after certain point (45%), the success rate decreases due to the negative noise impact of background traffic. When background traffic is very high (50%), the success rate becomes higher with collision than frame count and busy time. It is because the abnormal characteristics of collision persists longer than frame count and busy time.

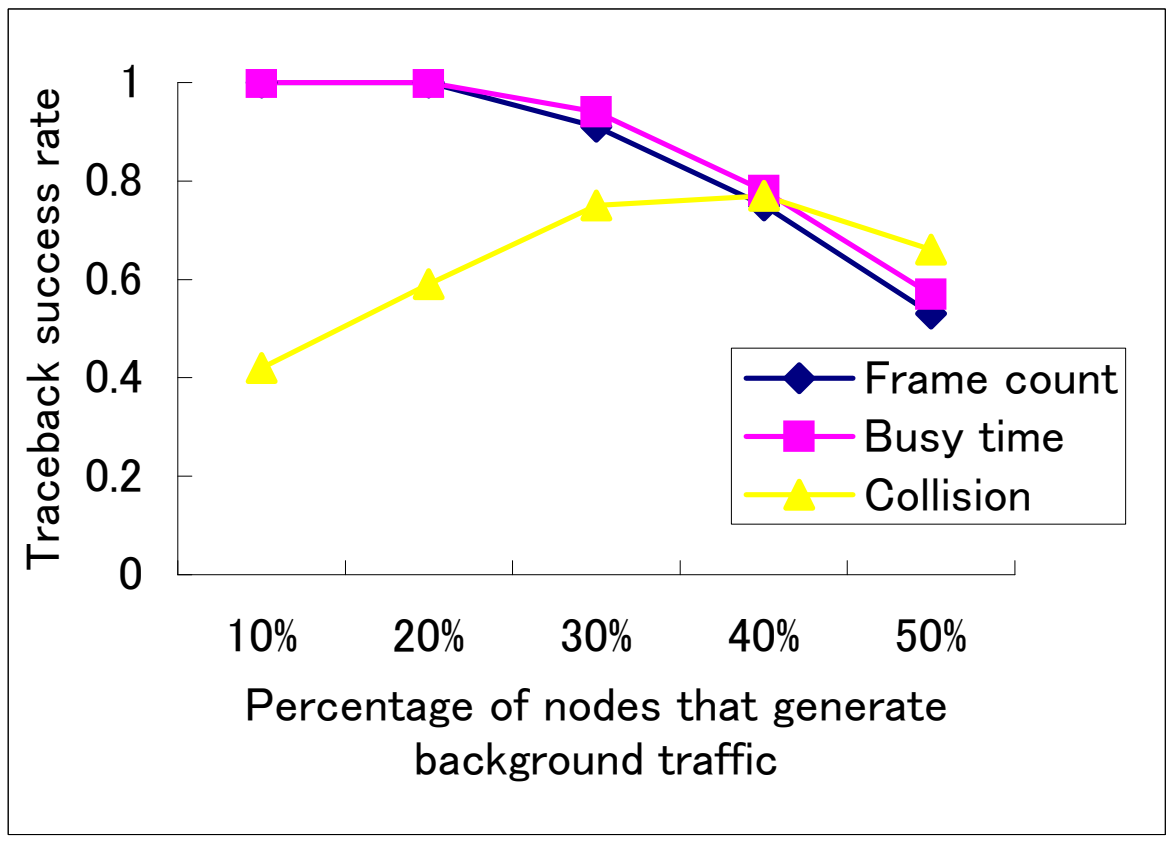

**[Figure 10] Traceback success rate**

We compared the traceback success rate between ATTENTION (frame count) and SWAT [13], which is based on traffic pattern matching. In Fig.11 ATTENTION shows higher success rate compared to SWAT especially when background traffic is increased (Avg.13.2% difference). It is because SWAT uses correlation of traffic pattern for attack signature matching and traceback. On the other hand, ATTENTION uses KS test for attack signature matching. KS test is less sensitive to traffic fluctuation than correlation coefficient method. In addition, ATTENTION is able to capture abnormality more sensitively using Eq.11. However, the drawback of ATTENTION is that it causes higher false positive rate as shown in Fig.12.

Fig.13 shows the number of nodes around attack path that observes attack signature. General traceback schemes (e.g., PPM, iTrace, logging, SWAT, etc) rely only on the intermediate nodes that relayed the attack traffic for traceback. On the other hand, ATTENTION tracks down attacker utilizing the information from overhearing nodes around the attack path. When node density (*N*: average number of nodes within transmission range) increases, more nodes are able to overhear the attack signature, which drastically increases the robustness against node compromise.

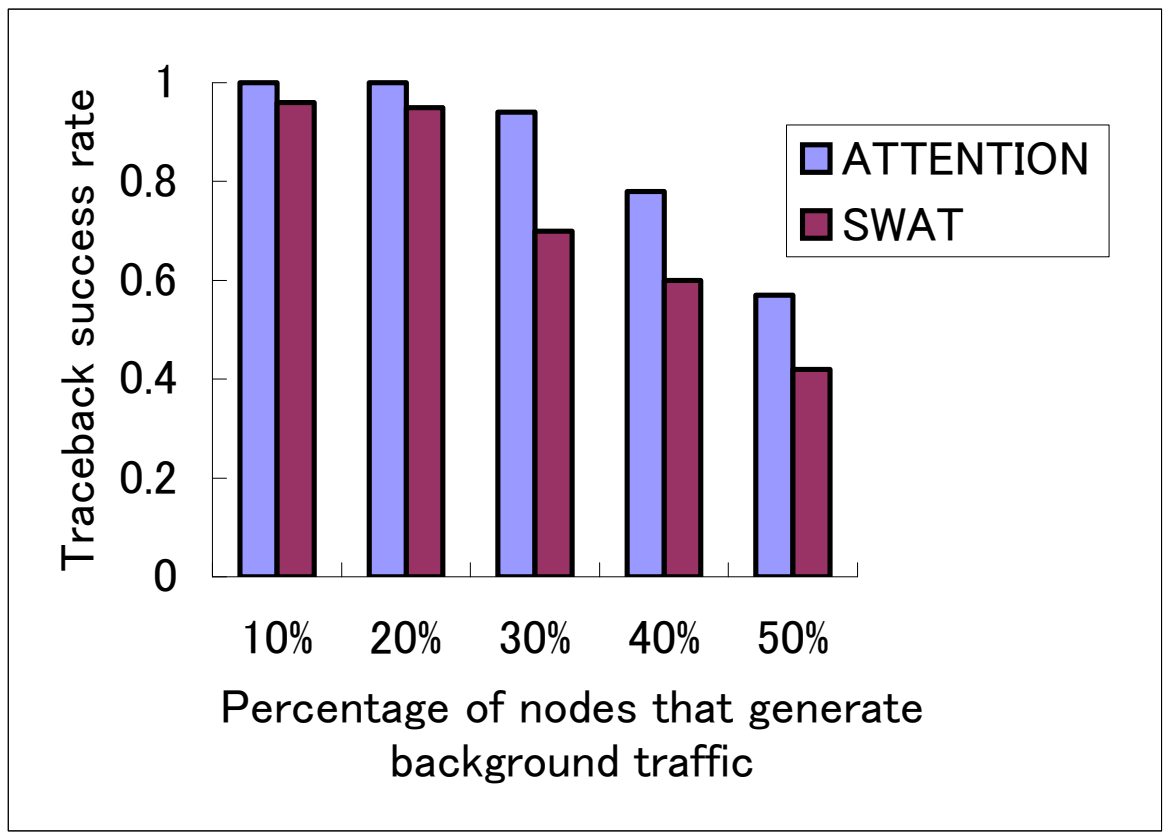

**[Figure 11] Traceback success rate comparison**

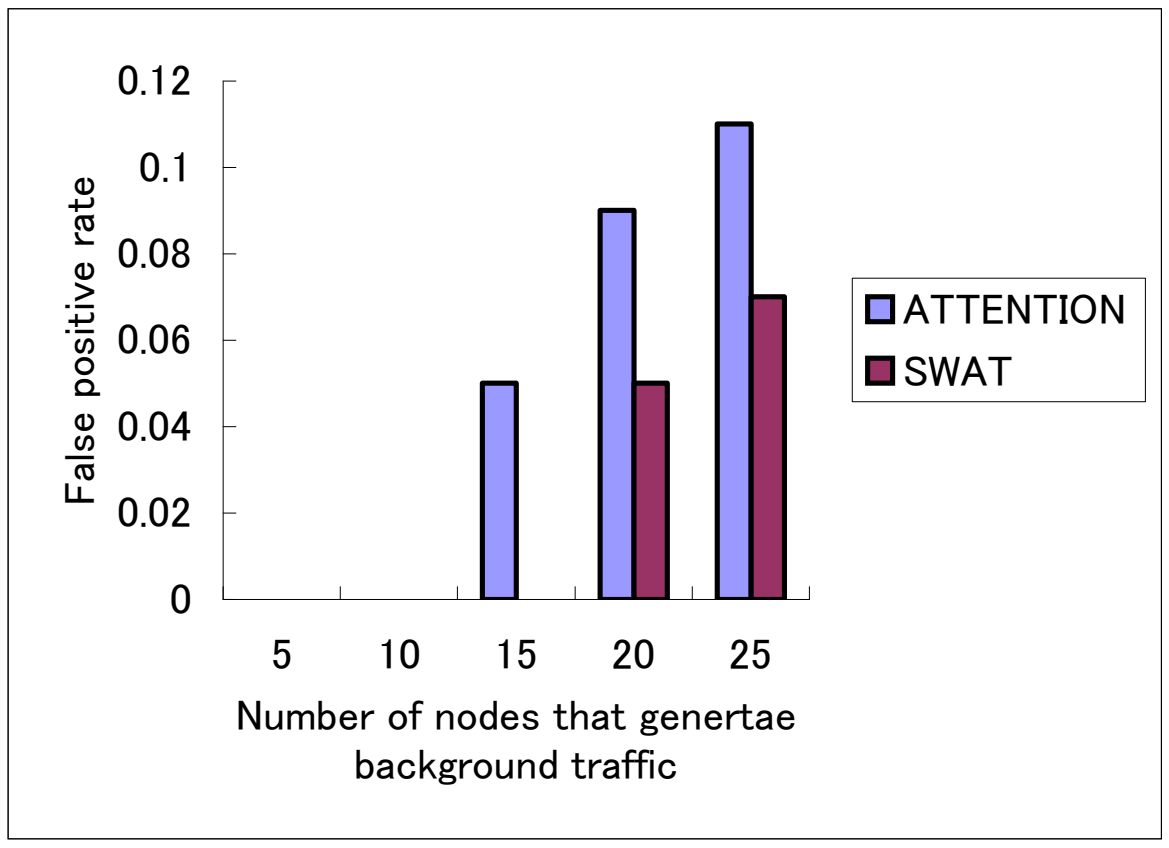

**[Figure 12] False positive rate comparison**

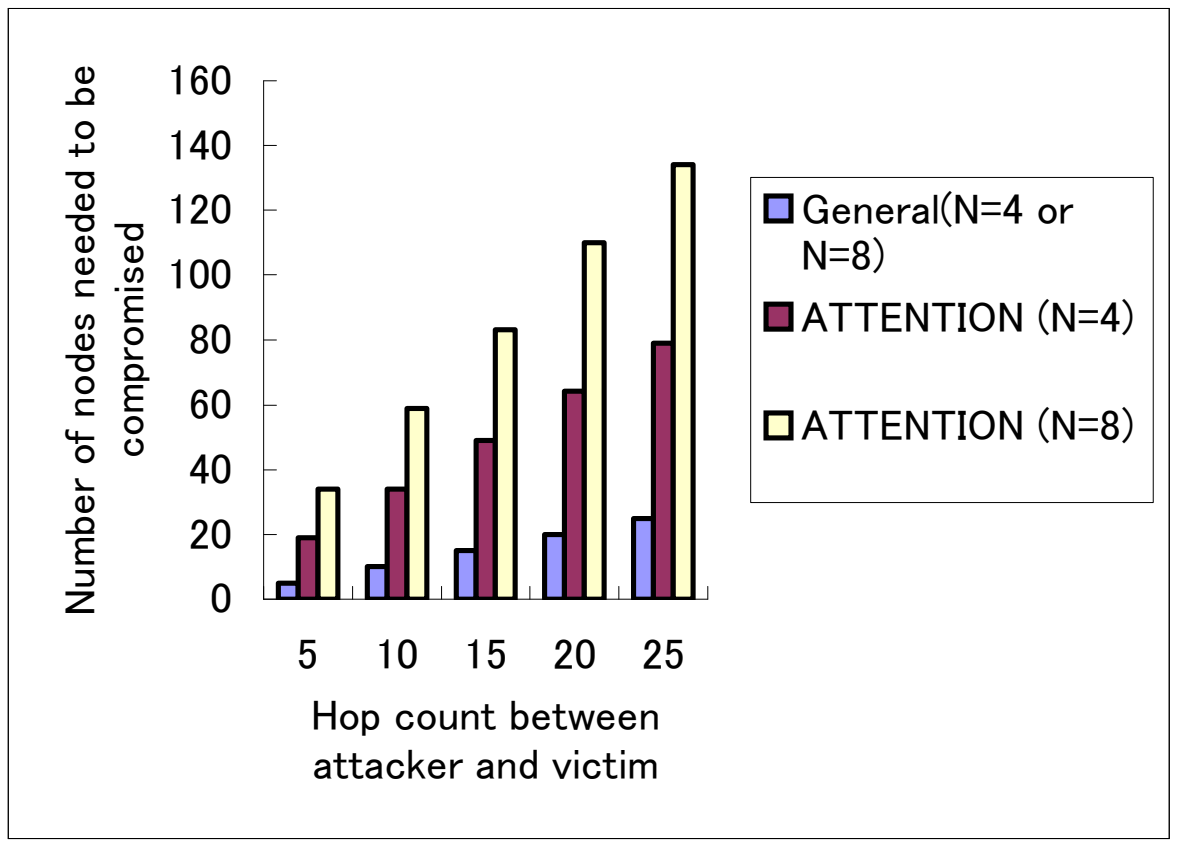

**[Figure 13] Robustness against node compromise**

## ■ Fine-grained traceback

Fig.14 shows DoS attacker traceback success rate in both ATTENTION Coarse-grained Traceback (ATTENTION-CT) scheme and ATTENTION Fine-grained Traceback (ATTENTION-FT) scheme. ATTENTION-FT shows high success rate (Avg.97%). The improvement becomes significant as background traffic is increased. It is due to the noise reduction impact as explained in Sec.5. Fig.15 shows DDoS attacker traceback success when there exist 10% of nodes that generate background traffic. SWAT shows very low success rate as the number of attackers is increased. It is because the abnormal characteristics of attack traffic is concealed by background traffic. ATTENTION-FT shows high success rate (Avg. 83%) since noise factors (by background traffic) are filtered out and abnormality of attack traffic can be effectively captured.

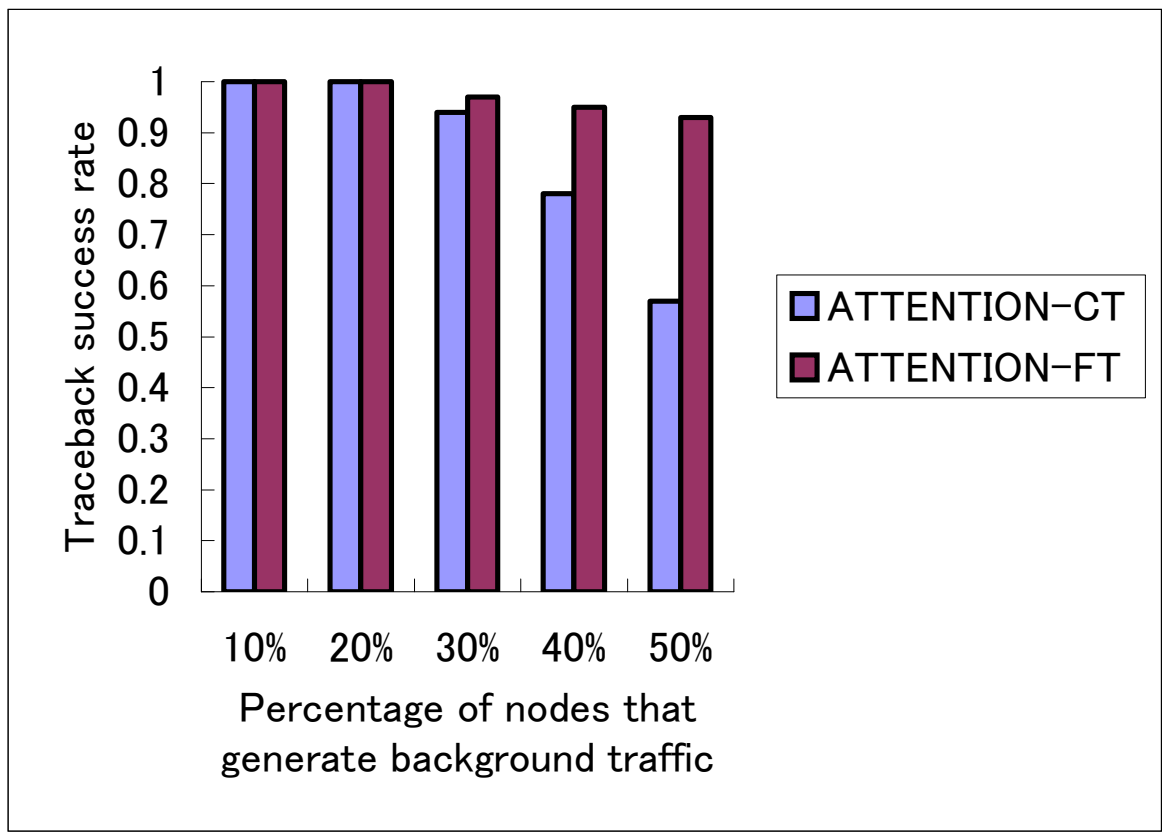

**[Figure 14] DoS attacker traceback success rate comparison**

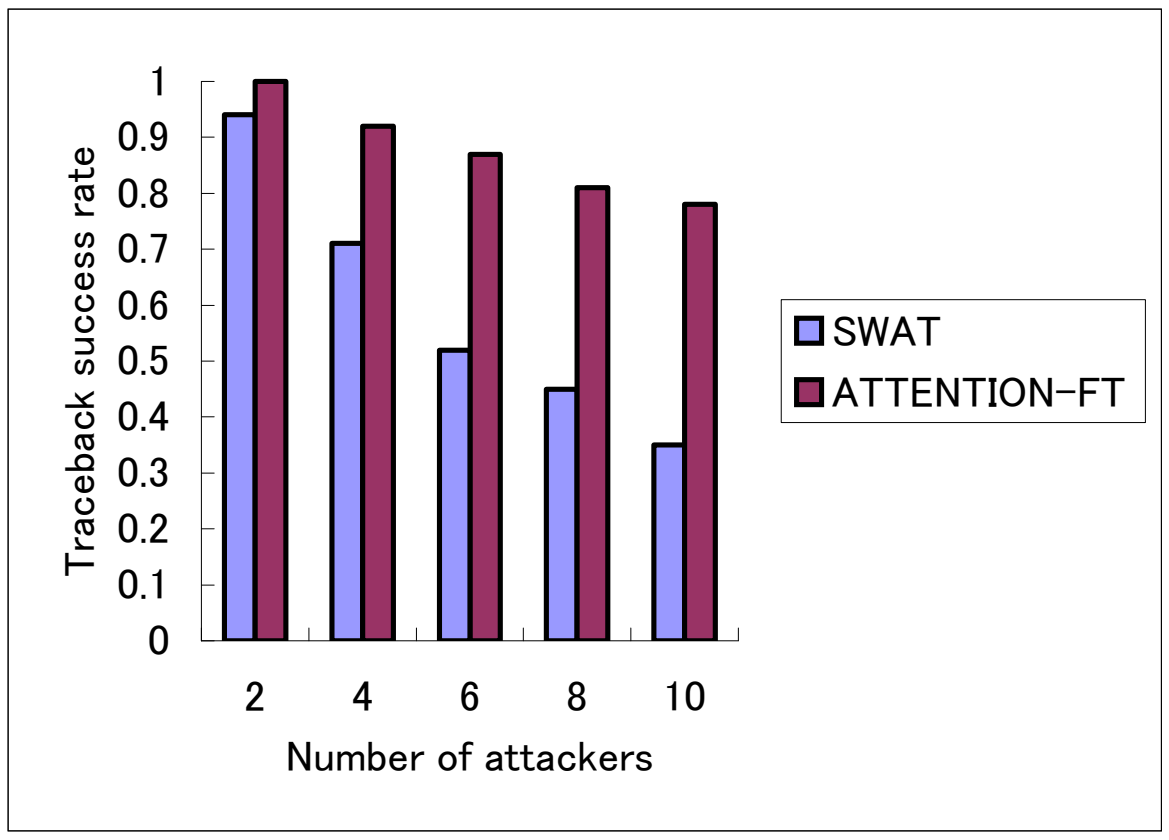

**[Figure 15] DDoS attacker traceback success rate comparison**

## ■ Overhead comparison

We compared communication overhead caused by traceback messages (number of packet transmission and reception). We compared ATTENTION with flooding-type search and ERS (Expanding Ring Search). As shown in Fig.16, ATTENTION incurs much less communication overhead than flooding and ERS. It is because query in ATTENTION is propagated only through the nodes that observe the attack signature. The overhead reduction becomes significant especially when the network size is large (74% reduction in 2500 nodes). Fig.17 shows similar overhead reduction for DDoS attacker traceback. Overhead reduction is low when network size is small and number of attackers is increased since ATTENTION reduces to flooding in such a case. However, when network size is large the overhead improvement becomes significant (41% reduction).

## ■ Mobile attacker traceback

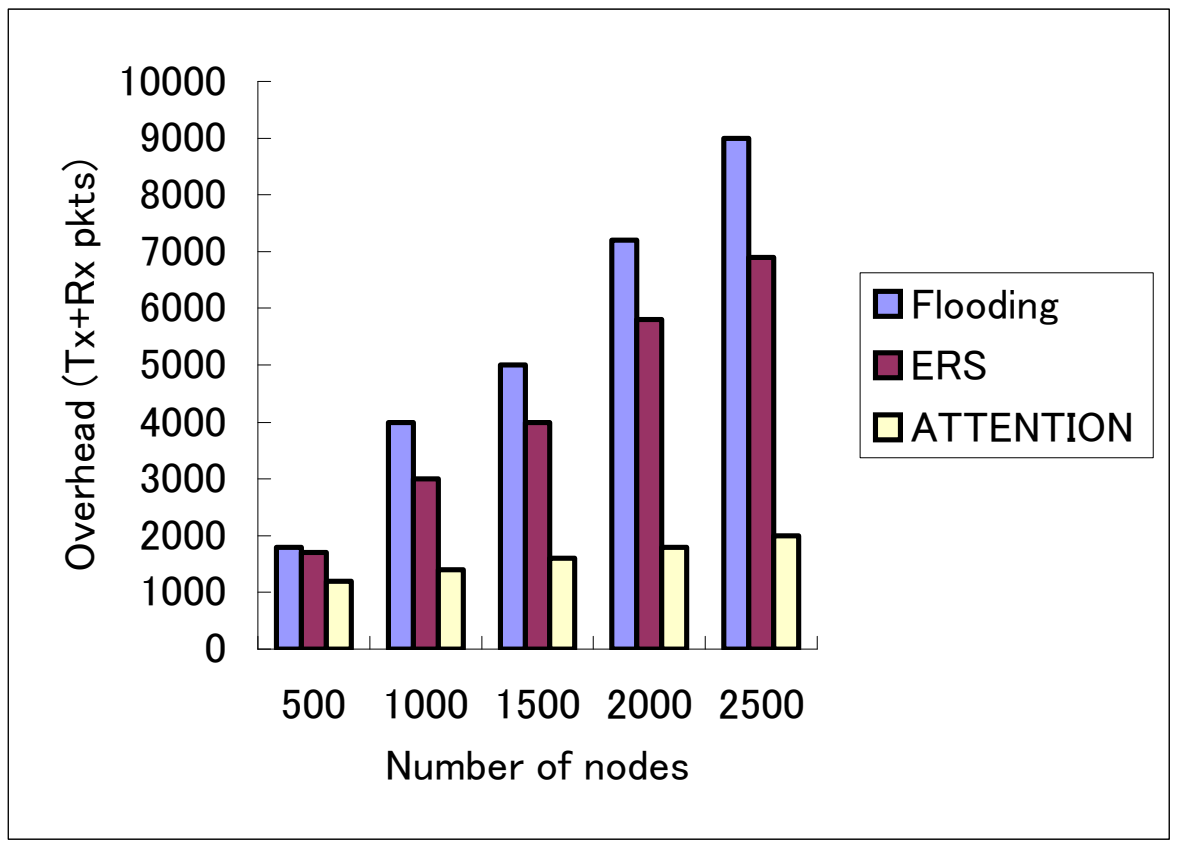

**[Figure 16] Overhead comparison for DoS attacker traceback**

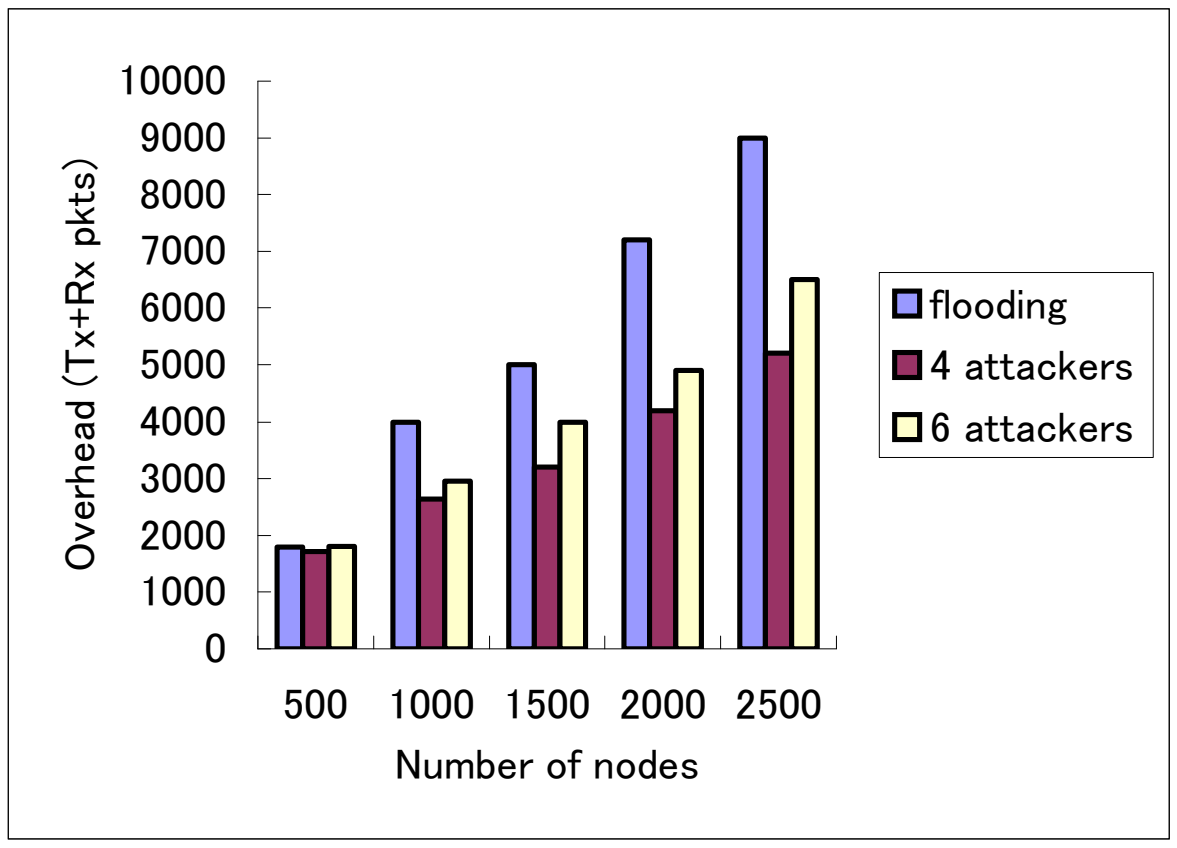

**[Figure 17] Overhead comparison for DDoS attacker traceback**

We compared SCF and TRF value between DDoS attack and mobile DoS attack. DDoS attack is performed from randomly selected 6 nodes. In mobile DoS attack, attacker and 5% of intermediate nodes are moving with random waypoint mobility model ($V_{max}$=2m/s, pause time=2.5s). Average SRF and TRF value are calculated where mobility is detected. We excluded the regions where α (Eq.16) is small (<0.1) since it implies that the nodes that reports attack signature moved out from original attack path. As shown in table 2, SRF is high in both in mobile attack and clustered DDoS attack since attack is observed in close region. SRF shows low value when DDoS attack is performed from geographically spread locations. TRF can differentiate between mobile attack and clustered DDoS attack since DDoS attack is launched around same time regardless of observation region. Consequently, we can effectively differentiate DDoS attack and mobile attack using combination SRF and TRF metrics (high TRF in mobile DoS attack and low TRF in clustered DDoS attack).

**[Table 2] Attack classification using SRF and TRF metrics**

| | SRF | TRF |
|---|---|---|
| Mobile DoS | 0.17 | 28.1 |
| Clustered DDoS | 0.18 | $5.12 \times 10^{-3}$ |
| Spread DDoS | 0.032 | $5.38 \times 10^{-3}$ |

- **Qualitative Comparison**

We provide qualitative comparison between ATTENTION and existing attacker traceback schemes in table 3. Existing IP traceback schemes are compared under the assumption that those are applied to MANETs.

## 8. DISCUSSION

- *Identification of real attackers*

Using ATTENTION, we can detect the next hop node to attack origin. However, ATTENTION is not able to detect the real identity of attackers if attackers disguise both their IP address and MAC address. In addition, the nodes that generate attack packets might be compromised innocent nodes. However, it is still important to trace back the closest points to the attack origins to take appropriate actions (e.g., filtering, isolation, etc).

- *Trust on contact and message integrity*

In our scheme, contacts are selected independently and randomly by each node to prevent divulgence of contact information and consequent compromise. Message between contacts-victim, contacts-contacts, and contact-vicinity nodes are relayed through intermediate nodes. Malicious intermediate node can inject false query/report or tamper the integrity of the message. To provide message integrity and authentication, we can leverage secure protocol such as [14]. However, it is out of scope of this paper.

- *MAC address spoofing of relay nodes and false report*

In coarse-grained traceback, ATTENTION is not susceptible to MAC address spoofing by relay nodes since we use regional abnormality and do not rely on specific MAC addresses for traceback. However, in fine-grained traceback, if relay node (not attacker) disguises its own MAC address, it can cause traceback failure. In this paper, we assumed that relay nodes only perform its given task (i.e., relay) and do not perform malicious activity.

Intermediate nodes can send false report to cause false positive and false negative in traceback. It can occur by inside attacker even if we use authentication and integrity protection schemes. However, our scheme can effectively prevent false positive/negative using majority voting. That is, even if some intermediate node does not report attack signature to cause false negative, we can still utilize information from other good relay/overhearing nodes that observes the attack signature. In addition, we ignore a report that does not agree with majority of reports, which prevent false positive.

- *Mobile attack detection*

For mobile attacker traceback, we performed limited simulation with simple scenario. As a future work, we will study and evaluate the impact of various mobility models [5] on traceabck success rate and overhead, etc. The mobility pattern (e.g., speed, direction, etc) of nodes may impact the success rate and efficiency of the proposed scheme.

## 9. CONCLUSION

We proposed attacker traceback protocol, ATTENTION, which utilizes MAC layer abnormality as

**[Table 3] Qualitative analysis of traceback schemes for MANETs**

| | Communication overhead | Computation (Analysis) overhead | Storage requirement | Mobile attacker traceability | DDoS attacker traceability |
|---|---|---|---|---|---|
| Controlled flooding [8] | High (short-term) | Low | N/A | No | Unable |
| Packet marking [17][18] | N/A | High at end-host | High at end-host | No | Poor |
| itrace[7] | Medium (long-term) | High at end-host | High at end-host | No | Poor |
| Logging[16] | Low (short-term) | High at intermediate node | High at Intermediate node | No | Good |
| SWAT[12][13] | Low (short-term) | Low | Low | No | Poor |
| ATTENTION-CT | Low (short-term) | Low | Low | No | Poor |
| ATTENTION-FT | Low (short-term) | Low | Medium | No | Good |
| ATTENTION *-Spatio-temporal fusion architecture* | Low (short-term) | Low | Low with CT/ Medium with FT | Yes | Good with FT |

attack signature. Under flooding type DoS/DDoS attack, MAC layer abnormality is observed around attack path from the attack origin to victim, which gives us a robust way for traceback. We verified that DoS attacker is successfully (Avg. 79%) using coarse-grained attack signature. In addition, with fine-grained attack signature, DoS attacker is traced with Avg.97% of success rate and DDoS attacker is traced with Avg. 83% of success rate. The communication overhead reduction is significant compared flooding-type search (74% in DoS, 41% in DDoS). We also proposed spatio-temporal fusion architecture to detect mobile attacker and verified that our architecture effectively distinguish DDoS attacker from mobile attacker and estimate current position of attacker using SRF and TRF metrics.

**[REFERENCES]**